\documentclass[apj,numberedappendix]{emulateapj}
\usepackage{apjfonts}

\shorttitle{QUASI-RELAXED STELLAR SYSTEMS IN TIDAL FIELD}
\shortauthors{VARRI \& BERTIN}
\submitted{Accepted for publication in The Astrophysical Journal}

\bibpunct[; ]{(}{)}{;}{a}{}{;}

\begin{document}

\title{Properties of quasi-relaxed stellar systems in an external tidal field}

\author{A. L. Varri and G. Bertin}
\affil{Universit\`{a} degli Studi di Milano, Dipartimento di Fisica, via Celoria 16, I-20133 Milano, Italy; anna.varri@unimi.it}

\begin{abstract}
 
In a previous paper, we have constructed a family of
self-consistent triaxial models of quasi-relaxed stellar systems,
shaped by the tidal field of the hosting galaxy, as an extension of
the well-known spherical King models. For a given tidal field, the
models are characterized by two physical scales (such as total
mass and central velocity dispersion) and two dimensionless
parameters (the concentration parameter and the tidal strength).
The most significant departure from spherical symmetry occurs when
the truncation radius of the corresponding spherical King model is
of the order of the tidal radius, which, for a given tidal strength,
is set by the maximum concentration value admitted. 
For such maximally extended (or ``critical") models the outer
boundary has a generally triaxial shape, given by the
zero-velocity surface of the relevant Jacobi integral, which is
basically independent of the concentration parameter. In turn, the
external tidal field can give rise to significant global
departures from spherical symmetry (as measured, for example, by
the quadrupole of the mass distribution of the stellar system)
only for low-concentration models, for which the allowed maximal
value of the tidal strength can be relatively high. In this paper
we describe in systematic detail the intrinsic and the projected
structure and kinematics of the models, covering the entire
parameter space, from the case of sub-critical (characterized by
``underfilling" of the relevant Roche volume) to that of critical
models. The intrinsic properties can be a useful starting point
for numerical simulations and other investigations that require
initialization of a stellar system in dynamical equilibrium. The
projected properties are a key step in the direction of a
comparison with observed globular clusters and other candidate
stellar systems.

\end{abstract}


\section{Introduction}

It is commonly thought that globular clusters can be described as
stellar systems of finite size, with a truncation in their density
distribution determined by the tidal field of the hosting galaxy.
Given the fact that many globular clusters undergo significant relaxation, 
the {  simplest analytical} models \citep{Kin66} {  that provide a 
successful description of these stellar systems} are defined by a 
quasi-Maxwellian distribution function $f_K(E)$, where $E$ is the 
single-star energy, with a truncation prescription continuous in phase 
space.

Most of the interesting physical mechanisms that underlie the
dynamical evolution of these stellar systems \citep[such as evaporation
and core collapse; e.g., see][]{Spi87,HegHut03} depend on such
truncation and are frequently studied in the context of
{\it spherical} models for which the action of tides
is implemented by means of the existence of a suitable truncation
radius (which, in principle, for a globular cluster of given mass
can be determined {\it a priori} if we know the cluster orbit and
the mass distribution of the hosting galaxy) {  supplemented by a 
recipe for the escape of stars}. {  Therefore, evolutionary models 
that rely on the assumption of spherical symmetry, such as Monte Carlo 
models \citep[e.g., for a description of two of the codes currently used, see][]
{Gie98,Jos00} and Fokker-Planck models \citep[e.g., for an example of, 
respectively, the isotropic and anisotropic case, see][]{CheWei90,TakLeeIna97} 
are necessarily based on an approximate treatment of the tidal field. In 
this context one interesting exception to the adoption of spherical symmetry 
is that of the axisymmetric Fokker-Planck models by \citet{EinSpu99}, 
which admit flattening induced by internal rotation.
}   

Yet, if tides are indeed responsible for the truncation, they
should also induce significant deviations from spherical symmetry:
{  in the simplest case of a cluster in circular orbit about 
the center of the hosting galaxy, the associated (steady) tidal field is 
nonspherical and determines an elongation of the mass distribution in 
the direction of the center of mass of the hosting galaxy accompanied 
by a compression in the direction perpendicular to the orbit plane  
\citep[e.g., see][]{Spi87,HegHut03}. Direct N-body simulations, in which 
an external tidal field can be taken into account explicitly, provide a 
unique tool for the study of the evolution of a tidally perturbed 
cluster, especially when non-circular orbits are considered so that 
tidal effects are time-dependent \citep[e.g., see][]{BauMak03}; 
in particular, this approach has led to detailed models of the rich 
morphology of tidal tails, i.e. streams of stars escaped from the cluster
\citep[e.g., see][]{LeeLeeSun06}.}

Indeed, deviations from spherical symmetry are observed in globular 
clusters {  \citep[e.g., see][]{Gey83,WhiSha87}}, but, they are 
often ascribed to other physical ingredients,
such as {  internal} rotation. In fact, it is recognized that the issue of what
determines the observed shapes of globular clusters remains
unclear \citep[e.g., see][and references therein]{Kin61,
Fall85,Han94,Ber08}.

One might argue that real globular clusters are likely to be not 
fully relaxed, may possess some rotation and experience time-dependent 
tides so that analytical refinements beyond the spherical one-component 
King models would not compete with the currently available numerical 
simulation tools (see \S~5.1 for additional comments) that allow us to 
include these and a great variety of other detailed effects that are 
relevant for the quasi-equilibrium configurations. 
However, physically simple analytical models, accompanied by the study of more 
realistic numerical simulations, serve as a useful tool to interpret 
real data and to provide insights into dynamical mechanisms, even though 
we know that real objects certainly include features that go well beyond 
such simple physical models. In this specific case, we argue that the 
triaxial geometry is a natural attribute of the physical picture of 
tidal truncation, which has already proved to provide useful guidance in 
the study of globular clusters. 

As demonstrated in a previous paper \citep[][hereafter Paper I]{BV08}, 
a complete characterization of the physically simple description of globular 
clusters as quasi-relaxed tidally-truncated stellar systems, in terms of 
fully self-consistent triaxial models, can be provided.
In order to better address the role of tides in determining the
observed structure of globular clusters, in Paper I
we have thus constructed self-consistent non-spherical equilibrium models of quasi-relaxed
stellar systems, obtained from the spherical case by including in
their distribution function the effects associated with the
presence of an external tidal field explicitly. We recall that our
models consider the stellar system in circular orbit within the
hosting galaxy, for simplicity assumed to be spherically-symmetric.
Therefore, in the corotating frame of reference, the tidal field
experienced by the system is static and the Jacobi integral $H$ is
available. In this physical picture the typical dynamical 
time associated with the orbits inside the cluster is assumed to be much 
smaller than the external orbital time. The procedure described in the previous paper starts
by replacing the single-star energy $E$ with the Jacobi integral
in the relevant distribution function $f_K(H)= A [\exp(-aH)-
\exp(-aH_0)]~$ if $H \le H_0$, with $H_0$ the cut-off constant,
and $f_K(H) = 0$ otherwise. Thus the collisionless Boltzmann
equation is satisfied. The construction of the self-consistent
models then requires the solution of the associated
Poisson-Laplace equation, that is of a second-order elliptic
partial differential equation in a free boundary problem, because
the boundary of the configuration, which represents the separation
between the Poisson and Laplace domains, can be determined only
{\em a posteriori}. The idea of using the Jacobi 
integral for the construction of tidal triaxial models had been proposed 
also by \citet{Wei93}. A family of models based on the distribution
function $f_K(H)$, but constructed with a different method, was 
discussed by \citet{HegRam95}; a comparison with our models will be 
given in \S~3.2.

Here we describe the properties of the resulting two-parameter
family of physically justified triaxial models, constructed with the
method of matched asymptotic expansions (see \S~4 and \S~A in
Paper I), inspired by previous studies of the formally similar
problem of rotating polytropes \citep{Smi75,Smi76}. In particular, we 
illustrate the properties of the first and second-order solutions.    

The paper is organized as follows. In \S~2 we present a thorough
description of the relevant parameter space. The intrinsic and
projected density distributions are discussed in \S~3, with
special emphasis on the global and local quantities that can be
used as diagnostics of deviations from spherical symmetry.
Intrinsic and projected kinematics are addressed in \S~4. The
concluding \S~5 gives the summary of the paper with a discussion
of the results obtained and a comment on the complex physical
phenomena that a large body of evolutionary models based on numerical 
investigations has shown to characterize the periphery of globular 
clusters.

\begin{figure}[t!]
\epsscale{1.16} \plotone{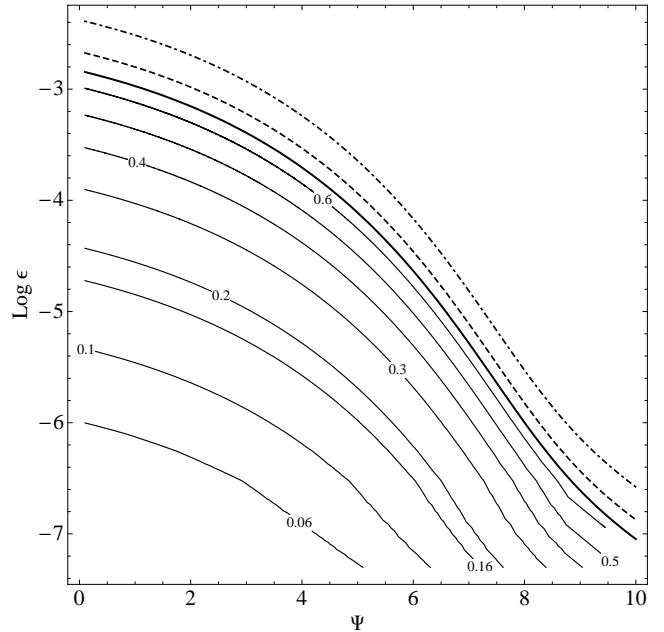} \caption{Parameter space for
second-order models. The uppermost solid line represents the
critical values of the tidal strength parameter for models in
which the potential of the hosting galaxy is Keplerian ($\nu=3$);
thin solid lines represent contour levels of the extension
parameter $\delta$. The dashed line is the critical curve for
models within a logarithmic potential ($\nu=2$). The dot-dashed
line gives the critical condition for a potential with $\nu=1$
(e.g., that of a Plummer sphere evaluated at $R_0=b/\sqrt{2}$,
with $b$ the model scale radius).} \label{parsp}
\end{figure}

\section{The Parameter Space}
 
The triaxial tidal models are characterized by two physical scales
(corresponding to the two free constants $A$ and $a$ in the
distribution function $f_K(H)$) and two dimensionless parameters.
The latter parameters are best introduced by referring to the
formulation of the Poisson equation in terms of the dimensionless
escape energy
\begin{equation}\label{psi}
\psi({  { \hat{r}}})=a\,H_0-[a\,\Phi_C({\it   \hat{r}})+\epsilon\,
T(\hat{x},\hat{z})]~,
\end{equation}
\noindent where $a\Phi_C$ is the {  dimensionless} cluster 
mean-field potential (to be determined self-consistently) and $T(\hat{x},\hat{z}) =
9(\hat{z}^2-\nu \hat{x}^2)/2$ represents the tidal potential (with the 
numerical coefficient $\nu = 4-\kappa^2/\Omega^2$, {  where $\kappa$ 
and $\Omega$ are respectively the epicyclic and orbital frequency}, 
depending on the potential of the hosting galaxy); the hat on the
spatial coordinates denotes that they are measured in units of the
scale length $r_0 = [9/(4 \pi G \rho_0 a)]^{1/2}$.
 
Then the first parameter, already available in the spherical case,
is the concentration of the system and can be expressed as a
dimensionless measure of the central depth of the potential well:
$\Psi\equiv \psi({  0})$. The second parameter, the tidal
strength $\epsilon$ in Eq.~(\ref{psi}), is defined as
\begin{equation}\label{eps}
\epsilon\equiv \frac{\Omega^2}{4 \pi G \rho_0}~,
\end{equation}
\noindent i.e. as the ratio of the square of the orbital frequency
(of the revolution of the stellar system around the center of the
hosting galaxy) to the square of the dynamical frequency associated
with the central density $\rho_0$ of the stellar system.
Alternatively, the effect of the tidal field can be measured by
the extension parameter
\begin{equation}\label{del}
\delta \equiv \hat{r}_{tr}/\hat{r}_{T}~,
\end{equation}
where $\hat{r}_{tr}=\hat{r}_{tr}(\Psi)$ is the {\em truncation} radius of the
spherical King model characterized by the same value of $\Psi$ and
$\hat{r}_T$ is the {\em tidal} (or {\em Jacobi}) radius, i.e. the
distance from the origin (the center of the stellar system) of the
two nearby Lagrangian points of the restricted three-body problem
considered in our simple physical picture. A given model will be
labelled by the pair of values $(\Psi,\epsilon)$ or, equivalently,
by the pair $(\Psi,\delta)$. {  The dimensionless cut-off constant $aH_0$ 
can be expressed in a natural way as an asymptotic series with respect to 
the tidal parameter $aH_0=\alpha_0+\alpha_1\epsilon+\alpha_2\epsilon^2/2+...$  
where the terms $\alpha_i$, as discussed in Paper I, depend only on $\Psi$.}
 
Much like the Hill surfaces for the standard restricted three-body
problem, we now consider the family of zero-velocity surfaces
defined by the condition $\psi({  \hat{r}})=0$, which represents
the boundary of our models. These surfaces can be open or closed,
depending on the value of the cut-off constant $a H_0$, which is
determined by the selected values of the two dimensionless
parameters that characterize the model. To be consistent with the
hypothesis of stationarity, we only consider closed
configurations. We call ``critical models" those that are bounded
by the critical zero-velocity surface (which is the outermost
available closed surface). For each value of $\Psi$, the critical
value of the tidal parameter can be found by (numerically) solving
the system
\begin{equation}
\left\{
\begin{array}{ll}\label{sistcrit}
\partial_{\hat{x}} \psi(\hat{x}=\hat{r}_T,\hat{y}=0,\hat{z}=0;\epsilon_{cr})=0\vspace{.1cm}\\
\psi(\hat{x}=\hat{r}_T,\hat{y}=0,\hat{z}=0;\epsilon_{cr})=0~,
\end{array}
\right.
\end{equation}
{  where the unknowns are $\hat{r}_T$ and $\epsilon_{cr}$}.
The method of matched asymptotic expansions proposed in Paper I for the 
solution of the relevant Poisson-Laplace equation requires an expansion 
in spherical harmonics, therefore it can be easily recognized that the 
first condition of Eq.~(\ref{sistcrit}) is equivalent to the requirement 
of vanishing gradient, which identifies the saddle points of the critical 
surface. In the general case, the condition $\partial_{\hat{x}} 
\psi(\hat{r}_T,0,0;\epsilon)=0$ determines the value 
of $\hat{r}_T$ for a given tidal strength $\epsilon${ , therefore 
$\hat{r}_T=\hat{r}_T(\Psi,\epsilon)$.}

By using in the escape energy defined in Eq.~(\ref{psi})
the zeroth-order expression for the cluster potential $(a\Phi_C^{\mbox{\tiny (ext)}})
^{(0)}(\hat{r})=\lambda_0/\hat{r}$ and for the cut-off constant\footnote{  We recall from Paper I 
that $\lambda_0=\hat{r}_{tr}^2{\psi_0^{\mbox{\tiny (int)}}}{'}(\hat{r}_{tr})$ and 
$\alpha_0=\lambda_0/\hat{r}_{tr}$, with $\psi_0^{\mbox{\tiny (int)}}$ the 
zeroth-order term of the asymptotic series of the internal solution. 
 $aH_0=\alpha_0$}, the system in Eq.~(\ref{sistcrit}) becomes
\begin{equation}
\left\{
\begin{array}{ll}\label{sistcrit0}
\displaystyle
\frac{\lambda_0}{\hat{r}_T^2}+9\,\epsilon_{cr}\,\nu\hat{r}_T=0\vspace{.1cm}\\
\displaystyle \alpha_0
-\frac{\lambda_0}{\hat{r}_T}+\frac{9}{2}\,\epsilon_{cr} \,\nu
\hat{r}_T^2=0 ~,
\end{array}
\right.
\end{equation}
thus leading to an expression for $\hat{r}_T^{(0)}$ in terms of
the dimensionless truncation radius of the corresponding spherical King 
model and a first estimate of the critical value of $\epsilon$
\begin{equation}\label{rT0}
\hat{r}_T^{(0)}= \frac{3}{2}\frac{\lambda_0}{\alpha_0}=
\frac{3}{2}\hat{r}_{tr}~,
\end{equation}
\begin{equation}\label{eps0}
\epsilon_{cr}^{(0)}=-\frac{8 \alpha_0^3}{243\,\nu\,\lambda_0^2}=
-\frac{8}{243\,\nu}\frac{1}{\hat{r}_{tr}^3}\lambda_0 ~.
\end{equation}
The first expression can also be written as
$\delta_{cr}^{(0)}=2/3$ \citep[see][]{Spi87}. We recall that 
{  $\lambda_0=\lambda_0(\hat{r}_{tr})$ and that $\hat{r}_{tr}=\hat{r}
_{tr}(\Psi)$}. Therefore, the right-hand side of Eqs.~(\ref{rT0}) and 
(\ref{eps0}) depends only on the value of $\Psi$. 

\begin{figure}[t!]
\vspace{0.1cm}
\epsscale{1.16}\plotone{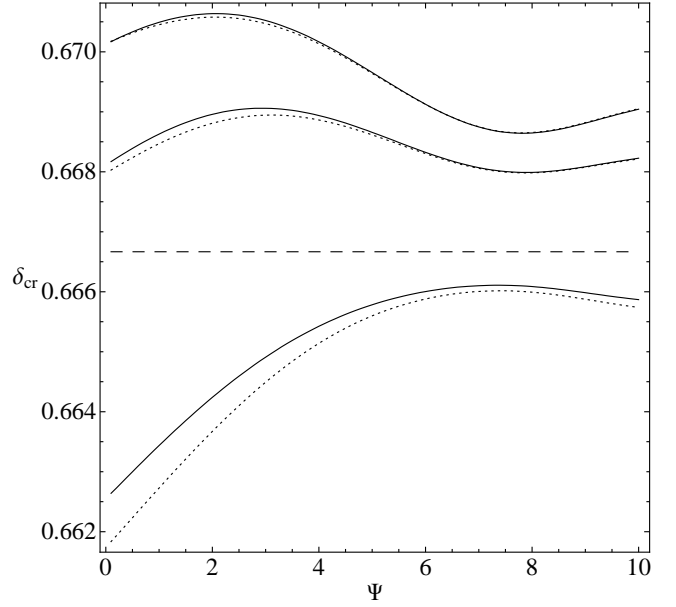} 
\caption{Critical values of the
extension parameter for first-order models (dotted) and
second-order models (solid) with different potentials of the
hosting galaxy ($\nu=3,2,1$ from top to bottom). The dashed
horizontal line shows the value 2/3 that is found when a zeroth-order 
approximation for $a\Phi_C$ is used (see Eq.~(\ref{rT0})).}\vspace{0.2cm} \label{delta}
\end{figure}

Here one important comment is in order. Strictly speaking, the
complete solution for $a\Phi_C$ derived by the method of matched
asymptotic expansions in \S~4 of Paper I is a well-justified
global uniform solution only for sub-critical (underfilled)
models. Close to the condition of criticality, i.e., when
$\hat{r}_{tr} \sim \hat{r}_{T}$, in the vicinity of the boundary surface 
the tidal term $\epsilon\,T$ (which is considered a small
correction in the construction of the asymptotic solution) becomes
comparable to the cluster term $a\Phi_C$, so that the asymptotic
solution is expected to break down. For such models, the iteration
method described in \S~5.2 of Paper I, which does not rely on the
assumption that the tidal term is small, is preferred and expected
to lead quickly to more accurate solutions. In practice, in line
with previous work on the similar problem for rotating polytropes
mentioned in the Introduction, we argue that the use of the
second-order asymptotic solution constructed in Paper I will give
sufficiently accurate solutions in the determination of the
critical value of the tidal parameter $\epsilon_{cr}$ from the
system in Eq.~(\ref{sistcrit}) and in the consequent assessment of
the general properties of models even when close to the critical
case. The main reason at the basis of this argument is that, even
for close-to-critical models, only very few stars populate the
region where the asymptotic analysis breaks down, so that the
overall solution should be only little affected. A direct
comparison between selected critical models calculated with both
the perturbation and the iteration method is presented in Appendix
A.
  
If we make use of the full second-order asymptotic solution for the 
escape energy {  $\left(\psi^{\mbox{\tiny (ext)}}\right)^{(2)}$, in 
which the second-order expressions for the cluster potential (recorded 
in Eq.~(\ref{multip})) and the cut-off constant are used}, the system 
in Eq.~(\ref{sistcrit}) can be re-arranged and written in standard form
\begin{equation}
\left\{
\begin{array}{ll}\label{sistcrit1}
A(\hat{r}_T)\epsilon_{cr}^2+2\,B(\hat{r}_T)\epsilon_{cr}+C(\hat{r}_T)=0  \vspace{.1cm}\\
D(\hat{r}_T)\epsilon_{cr}^2+2\,E(\hat{r}_T)\epsilon_{cr}+F(\hat{r}_T)=0~,
\end{array}
\right.
\end{equation}
with
\begin{eqnarray}\label{coeffA}
& \displaystyle A(\hat{r}_T)=\lambda_2\hat{r}_T^4-(b_{20}-b_{22}\sqrt{3})\frac{3}{4}
\sqrt{\frac{5}{\pi}}\hat{r}_T^2+b_{40}\frac{45}{16\sqrt{\pi}}&\nonumber \\
& \displaystyle -b_{42}\frac{15}{8}\sqrt{\frac{5}{\pi}}+b_{44}\frac{15}{16}\sqrt{\frac{35}{\pi}}~,&
\end{eqnarray}
\begin{equation}\label{coeffB}
B(\hat{r}_T)=\lambda_1\hat{r}_T^4-(a_{20}-a_{22}\sqrt{3})\frac{3}{4}
\sqrt{\frac{5}{\pi}}\hat{r}_T^2+9\nu\hat{r}_T^7~,
\end{equation}
\begin{equation}\label{coeffC}
C(\hat{r}_T)=2\lambda_0\hat{r}_T^4~,
\end{equation}
\begin{eqnarray}\label{coeffD}
& \displaystyle D(\hat{r}_T)=\alpha_2\hat{r}_T^5-\lambda_2\hat{r}_T^4+(b_{20}-b_{22}\sqrt{3})
\frac{1}{4}\sqrt{\frac{5}{\pi}}\hat{r}_T^2 \nonumber \\
& \displaystyle -b_{40}\frac{9}{16\sqrt{\pi}}
+b_{42}\frac{3}{8}\sqrt{\frac{5}{\pi}}-b_{44}\frac{3}{16}\sqrt{\frac{35}{\pi}}~,&
\end{eqnarray}
\begin{equation}\label{coeffE}
E(\hat{r}_T)=\alpha_1\hat{r}_T^5-\lambda_1\hat{r}_T^4+(a_{20}-a_{22}\sqrt{3})
\frac{1}{4}\sqrt{\frac{5}{\pi}}\hat{r}_T^2+\frac{9}{2}\nu\hat{r}_T^7~,
\end{equation}
\begin{equation}\label{coeffF}
F(\hat{r}_T)=2(\alpha_0\hat{r}_T^5-\lambda_0\hat{r}_T^4)~,
\end{equation}
where the relevant constants ($\lambda_i$ with $i=0,1,2$, $a_{lm}$ with 
$l=0,2$ and $b_{lm}$ with $l=0,2,4$ $m=0,2,...,l$), which are determined by the matching
process, are defined in \S~4.4 of Paper I. [The corresponding
system based on the first-order solution for $\psi^{\mbox{\tiny (ext)}}$ can be
recovered by setting $A(\hat{r}_T)= D(\hat{r}_T) = 0$.] This
system has been solved numerically, by means of the Newton-Raphson
method, since the equations are nonlinear in $\hat{r}_T$ (in
particular, they are polynomials of fifth and seventh-order, for
the first and second-order solution respectively). As noted in the
discussion of the simpler Eq.~(\ref{sistcrit0}), the solution for
$\epsilon_{cr}$ can then be represented as a function of the
concentration parameter $\Psi$. The parameter space of the first
and second-order models has been explored by means of an equally
spaced grid from $\epsilon=5 \times 10^{-8}$ to $\epsilon=1 \times
10^{-2}$ at steps of $5 \times 10^{-7}$ and from $\Psi=0.1$ to
$\Psi=10$ at steps of $0.1$.
  
The parameter space for the second-order models is presented in
Fig.~\ref{parsp}. The plot provides the contour levels of the
extension parameter $\delta$, with the uppermost solid line
corresponding to $\delta = \delta_{cr} \approx 2/3$ (thus
identifying the critical models), based on the choice $\nu = 3$
(Keplerian host galaxy). The critical curves for $\nu = 2$ (host
galaxy characterized by flat rotation curve) and for $\nu = 1$
(Plummer potential evaluated at $R_0=b/\sqrt{2}$, with $b$ the
model scale radius) are shown as a dashed and dot-dashed line,
respectively.
 
Sub-critical, underfilled models (bottom-left corner of the
figure), with $\delta \ll \delta_{cr}$, are only little
affected by the tidal perturbation. The maximally deformed models
are those with $\delta \approx \delta_{cr}$ (close to the
uppermost solid line, i.e. close-to-critical configurations).
Figure~\ref{parsp} shows that the critical value for the tidal
strength parameter depends strongly on concentration, with a
variation of almost four orders of magnitude in the explored range
of $\Psi$. The figure also indicates that for lower values of
$\nu$ the critical curve moves upwards, i.e. the available
parameter space increases.
 
The difference between the critical value of the tidal strength
parameter for first and second-order models (for a chosen value of
$\nu$) is very small, around $10^{-5}$ for low-concentration models,
down to $10^{-9}$ or less for models with $\Psi \approx 10$. 
The critical value of the extension parameter $\delta$ depends only 
weakly on concentration and on $\nu$, as illustrated in Fig.~\ref{delta}. 

In closing this section, se should reiterate that, in spite of the 
abundant use of symbols required by the analysis, the family of models 
that we have studied is characterized by two dimensionless parameters 
$(\Psi,\epsilon)$. [As an alternative pair, we may refer to the standard 
concentration parameter $C=log(r_{tr}/r_0)$, equivalent to $\Psi$ and 
frequently used in the context of spherical King models, and to the 
extension parameter $\delta=\hat{r}_{tr}/\hat{r}_T$, equivalent to 
$\epsilon$.] The free constants $A$ and $a$ that appear in 
the distribution function $f_K(H)$ set the two physical scales. In turn, 
the models constructed by \citet{HegRam95} are a one-parameter family of 
models, because these authors focused on the critical case and did not 
discuss the sub-critical regime.
\begin{figure}[t!]
\epsscale{1.25} \plotone{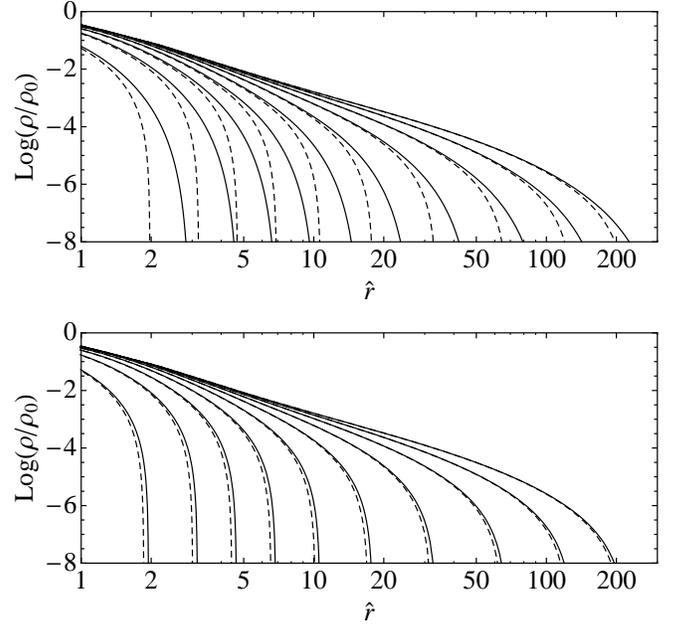} 
\caption{Intrinsic density
profiles (normalized to the central value) for critical
second-order models with $\nu=3$ and $\Psi=1,2,...,10$ (from left
to right). Top panel (a): profile of the triaxial models along the
$\hat{x}$-axis (solid) and of the corresponding spherical King
models (dashed). Bottom panel (b): profile of the triaxial models
along the $\hat{y}$-axis (solid) and the $\hat{z}$-axis (dashed).}
\label{dens}
\end{figure}

\section{Intrinsic and projected density distribution}
 
\subsection{Intrinsic density profile}
 
The models are characterized by reflection symmetry with respect
to the three natural coordinate planes. With respect to the
unperturbed configuration (i.e. the spherical King model with the
same value of $\Psi$), they exhibit an elongation along the
$\hat{x}$-axis (defined by the direction of the center of the host
galaxy), a compression along the $\hat{z}$-axis (the direction
perpendicular to the orbit plane of the globular cluster), and
only a very modest compression along the $\hat{y}$-axis.
 
Models with $\delta \le 0.2$, regardless of the value of $\Psi$,
are practically indistinguishable from the corresponding spherical
King models; significant departures from spherical symmetry occur
for models with $\delta \approx 0.4$ or higher. In
Fig.~\ref{dens}.a we show the density profile along the
$\hat{x}$-axis for a selection of critical second-order models
with $\nu=3$ in comparison with that of the corresponding
spherical King models; note that for a model with $\Psi=2$ the
elongation is already significant at $Log(\rho/\rho_0)\approx-4$,
while for a model with $\Psi=8$ a similar elongation is reached
only at much lower density levels ($Log(\rho/\rho_0)\approx-7$).
The corresponding profiles along the $\hat{y}$-axis and the
$\hat{z}$-axis are given in Fig.~\ref{dens}.b.

For completeness, we checked the dependence of our density
profiles on the potential of the host galaxy. Consistent with the
general trends suggested by Fig.~\ref{delta}, the elongation
along the $\hat{x}$-axis and the compression along the
$\hat{z}$-axis for the models with $\nu=3$ turn out to be slightly
weaker than for the models with smaller values of $\nu$.

\begin{figure}[t!]
\vspace{0.1cm}
\epsscale{1.15} \plotone{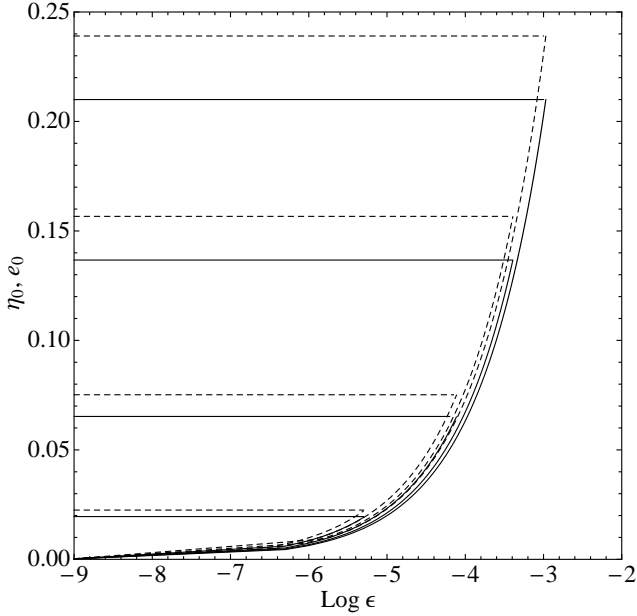} 
\caption{Central
values of the polar ($e_0$; dashed) and equatorial ($\eta_0$;
solid) eccentricities of the isodensity surfaces of second-order
models with $\nu=3$, $\Psi=1,3,5,7$ (from top to bottom) and
$\epsilon \in [0,\epsilon_{cr}(\Psi)]$. Horizontal lines
mark the maximum value of $e_0$ and $\eta_0$ reached by
the critical models (i.e., for $\epsilon=\epsilon_{cr}(\Psi)$).}\vspace{0.1cm}
\label{ecce0}
\end{figure}
  
Since the dimensionless density distribution of a model,
identified by $(\Psi,\epsilon)$, is given by
$\hat{\rho}=\hat{\rho}[\psi({  \hat{r}})]$, where $\psi$ is the
dimensionless escape energy and $\hat{\rho}$ is a monotonically
increasing function that vanishes for vanishing argument (see
Eq.~(11) in Paper I), there is a one-to-one correspondence between
isodensity and isovelocity surfaces, the latter being defined by
the condition $\psi^{(int)}({  \hat{r}})=S$, where $S$ is a
constant (with $0 \leq S \leq \Psi$). We recall that nonspherical 
models often exhibit equipotential surfaces rounder than the 
isodensity surfaces \citep[e.g., see][]{Eva93,CioBer05}. 
The reason for the presence of this property in our
models is that the supporting distribution function depends only on 
the Jacobi integral, i.e. the (isolating) energy integral in the 
rotating frame. Therefore, for each value of $S$, we can define 
the semi-axes $\hat{a}$, $\hat{b}$, and $\hat{c}$ of the corresponding triaxial 
isodensity surface, by means of the intersections of the surface 
with the $\hat{x}$, $\hat{y}$, and $\hat{z}$ axes, which turn out to 
follow the ordering $\hat{a}\ge\hat{b}\ge\hat{c}$. The shape of the 
triaxial configuration can thus be described in terms of the polar and
equatorial eccentricities, defined as
$e=[1-(\hat{c}/\hat{a})^2]^{1/2}$ and
$\eta=[1-(\hat{b}/\hat{a})^2]^{1/2}$, respectively.
 
A surprising result can be derived analytically. In the innermost
region $\hat{r}\ll\hat{r}_{tr}$ (i.e., for $S \sim \Psi$), the
dimensionless escape energy can be expanded to second order in the
dimensionless radius
\begin{eqnarray}\label{psiint}
& \displaystyle \psi^{(int)}({\bf \hat{r}}) \sim \Psi -\frac{3}{2}\hat{r}^2 +
\epsilon\hat{r}^2\left[-\frac{3}{2}(1-\nu)+A_{20}Y_{20}(\theta,\phi)
+ \right. \\
& \displaystyle \left. A_{22}Y_{22}(\theta,\phi)\right]+\frac{\epsilon^2}{2}\hat{r}^2\left[(1+B_{20}) Y_{20}(\theta,\phi)
+ (1+B_{22})Y_{22}(\theta,\phi)\right]~.& \nonumber 
\end{eqnarray}
Here some terms of the second-order solution do not contribute
(e.g., it can be readily checked that
$\psi_{2,4m}^{(int)}(\hat{r})\sim\hat{r}^4$ and
$\psi_{2,00}^{(int)}(\hat{r})\sim\hat{r}^6$). Then by setting
$\psi^{(int)}(\hat{a},0,0)=\psi^{(int)}(0,\hat{b},0)
=\psi^{(int)}(0,0,\hat{c})$, we find that in the innermost region
the eccentricities tend to the following non-vanishing central
values
\begin{equation}\label{e0}
e_0=\frac{\{\epsilon(A_{22}-\sqrt{3}A_{20}+(\epsilon/2)[1+B_{22}-\sqrt{3}
(1+B_{20})])\sqrt{15/\pi}\}^{1/2}}{\{6+2\epsilon[3(1-\nu)-A_{20}\sqrt{5/\pi}]
-\epsilon^2(1+B_{20})\sqrt{5/\pi}\}^{1/2}}~,
\end{equation}
\begin{equation}\label{eta0}
\eta_0=\frac{\{\epsilon[2A_{22}+(1+B_{22})\epsilon]\sqrt{15/\pi}\}^{1/2}}{\{6+\epsilon d_1+(\epsilon^2/2)d_2\}^{1/2}}~,
\end{equation}
where
\begin{equation}\label{d1}
d_1=6(1-\nu)+(A_{20}+\sqrt{3}A_{22})\sqrt{5/\pi}
\end{equation}
\begin{equation}\label{d2}
d_2=[1+B_{20}+\sqrt{3}(1+B_{22})]\sqrt{5/\pi}
\end{equation}
which depend explicitly on the tidal strength and implicitly on
the concentration. This result is nontrivial. In fact, since the
tidal potential is a homogeneous function of the spatial
coordinates, naively we might expect that in their central region
the models reduce to a perfectly spherical shape (i.e.,
$e_0=\eta_0=0$), even for finite values of the tidal strength.
Instead, $e_0$ and $\eta_0$ are $\mathcal{O}(\epsilon^{1/2})$ and
strictly vanish only in the limit of vanishing tidal strength.

\begin{figure}[t!]
\epsscale{1.11} \plotone{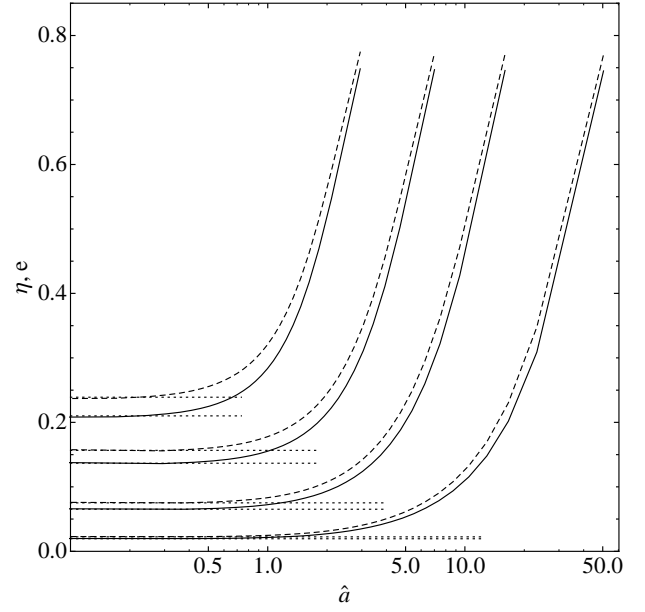} \caption{Profiles of
the polar ($e$; dashed) and equatorial ($\eta$; solid)
eccentricities of the isodensity surfaces for selected critical
second-order models with $\nu=3$ and $\Psi=1,3,5,7$ (from left to
right). Dotted horizontal lines show the central eccentricity
values (see Eqs.~(\ref{e0}) and (\ref{eta0})). }\vspace{0.1cm} \label{ecce}
\end{figure} 
 
Figure \ref{ecce0} shows the central values of the eccentricities
for second-order models with $\nu=3$ and selected values of
concentration, as a function of tidal strength within the range
$[0,\epsilon_{cr}(\Psi)]$. Consistent with the general trends
identified in the discussion of the parameter space,
low-concentration models show the most significant departures from
spherical symmetry. The full eccentricity profiles (as a function
of the major axis) are shown in Fig.~\ref{ecce} for a selection of
critical second-order models; here the calculation of $e$ and
$\eta$ has been performed by numerically determining the values of
the semi-axes of a number of reference isovelocity surfaces,
defined by $\psi^{(int)}({  \hat{r}})=S_i=(25-i)\Psi/25-0.01$
with $i=0,..,25$. Outside the central region, the profiles
increase monotonically and, independently of concentration, at the
boundary they reach approximately a fixed maximum value
($e\approx0.78$ and $\eta \approx0.74$), which corresponds to the
fact that the shape of the boundary surface of a critical model
($\delta_{cr}^{(0)}=2/3$; see also Fig.~\ref{delta}) depends only
modestly on concentration.
\vspace{0.8cm}
	
\subsection{Comparison with the models constructed by \citet{HegRam95}}

The method used in Paper I for the construction of the models 
illustrated in this paper can be summarized as follows. 
The solution in the internal (Poisson) and external (Laplace) domains are
expressed as an asymptotic series with respect to the dimensionless parameter 
$\epsilon$, representing the tidal strength (defined in Eq.~(\ref{eps})), 
which is considered to be small. The $k$th-order term of the asymptotic 
series of the internal and external solution are denoted by  
$\psi_k^{\mbox{\tiny (int)}}({  \hat{r}})$ and 
$\psi_k^{\mbox{\tiny (ext)}}({  \hat{r}})$ respectively, so that  
the zeroth-order terms define the standard spherical King models (see \S~4.1 
in Paper I for a detailed discussion). The quantity 
$(\psi^{\mbox{\tiny(ext)}})^{(k)}$ indicates the $k$th-order external solution, 
i.e. the corresponding asymptotic series truncated at the term 
$\psi^{\mbox{\tiny(ext)}}_k$; a similar notation holds for the internal solution. 
The validity of the expansion breaks down where the second term is comparable 
to the first, i.e. where $\psi_0=\mathcal{O}(\epsilon)$. This singularity 
is cured by introducing a {\em boundary layer} in which both the spatial 
coordinates and the solution $\psi^{\mbox{\tiny(lay)}}$ are suitably rescaled 
with respect to the tidal parameter. To obtain a uniformly valid solution 
over the entire space, an asymptotic matching \citep[see][Eq.~(5.24)]{Dyk75} 
is performed between the pairs $(\psi^{\mbox{\tiny(int)}},\psi^{\mbox{\tiny(lay)}})$ 
and $(\psi^{\mbox{\tiny(lay)}},\psi^{\mbox{\tiny(ext)}})$. Each term $\psi_k({  \hat{r}})$ 
is then expanded in spherical harmonics with radial coefficients 
$\psi_{k,\,lm}(\hat{r})$. The internal region requires a numerical solution 
of the Cauchy problems for the radial coefficients (we used a fourth-order 
Runge-Kutta code) while in the external region a formal solution with 
multipolar structure is available and in the boundary layer the integration 
in the radial variable can be performed analytically (see \S~4.2 and 4.3 
in Paper I, respectively).

\begin{figure}[t!]
\epsscale{1.15} \plotone{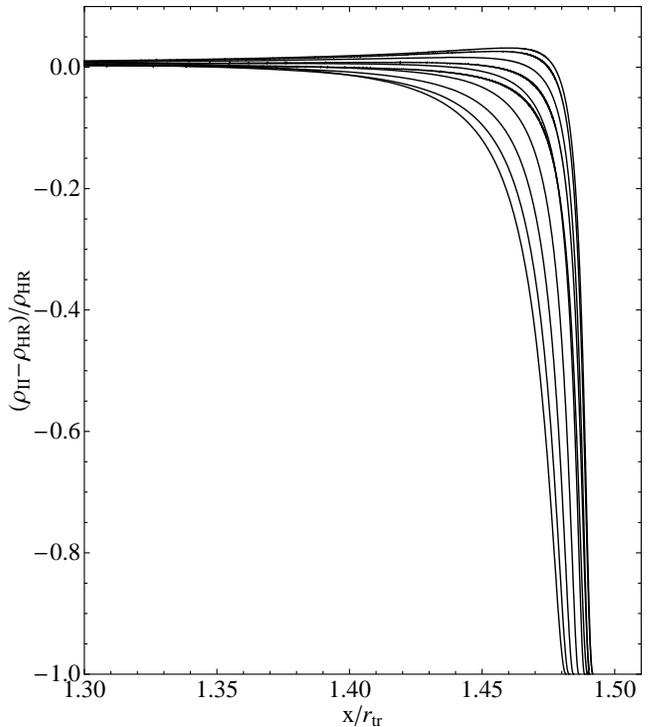} \caption{  Relative difference between 
the intrinsic density profiles of critical second-order models ($\rho_{II}$) 
constructed in this paper and those of the corresponding (first-order) models 
($\rho_{HR}$) described by \citet{HegRam95}. The comparison has been performed along 
the three axes in the whole internal+boundary region (see main text). Here 
we illustrate the difference along the $x$-axis in the boundary layer
($\Psi=1,2,...,10$ from left to right). At variance with Fig.~\ref{dens},
the spatial coordinate is scaled with respect to the truncation radius 
instead of the scale radius $r_0$.}
\label{conf}
\end{figure}

{  The models described by \citet{HegRam95} are also based on a 
perturbation approach, but the method used is different from ours, provides 
a solution of the Poisson equation that is first-order with respect to 
the tidal parameter, and is restricted to the ``critical" case; actually, 
as noted in \S~2 after Eq.~(\ref{eps0}), the perturbation approach is bound to break 
down in the critical case. Technically speaking, their method is in the 
form of a ``patching" procedure, in contrast with our asymptotic matching. 
Therefore, the models constructed in Paper I, while consistent, to first order,  
with those of \citet{HegRam95}, are more general. We also recall that our 
method is also applicable to systems described by different distribution 
functions (see Appendix B and C in Paper I).}

{  We have thus performed a quantitative comparison between the
intrinsic density profiles of our critical second-order models and those of 
the models\footnote{  For this purpose, we used the code that implements 
the models described by \citet{HegRam95}, written by D.~C. Heggie and 
available within the STARLAB software environment \citep[][]{Starlab}} 
by \citet{HegRam95}, both referred to the case in which the host galaxy 
is Keplerian ($\nu=3$). As desired, there is substantial consistency except 
for the outermost part of the boundary layer in which our models are 
slightly more compact, due to a global ``boxiness" effect induced by 
the second-order term present in our models in which harmonics of order 
$l=4$ also play a role. In Fig.~\ref{conf} we represent the relative difference 
between the two density profiles evaluated along the $x$-axis (with 
the coordinates scaled with respect to the truncation radius instead of 
the usual scale radius $r_0$) for selected values of $\Psi$. A similar 
behaviour is found also along the $y$-axis, for $y/r_{tr}$ in the range 
$[0.80,1]$, and along the $z$-axis, for $z/r_{tr}$ in the range 
$[0.80,0.95]$. In the central part of the internal region, along the 
principal axes, the relative difference is smaller than 5 percent for 
every value of $\Psi$ we tested, while near the transition to the boundary 
layer (i.e $x/r_{tr}\lesssim 1$ and $y/r_{tr},\,z/r_{tr}\lesssim 0.8$) a 
difference of 20 percent can be reached in the case of highly concentrated models 
($\Psi=8,9,10$). We interpret these differences as due to the combined 
effects of the patching vs. matching adopted process and of the different 
grid on which the Cauchy problems for the radial coefficients are solved 
(we used a regular radial grid while \citet{HegRam95} used a more complex 
tabulation resulting from their choice of taking the zeroth-order cluster 
potential as the independent variable and of the function $ln(1+\hat{r}^2)$ 
instead of $\hat{r}$).           
}

\begin{figure*}[t!]
\vspace{0.4cm}
\epsscale{1.1} \plotone{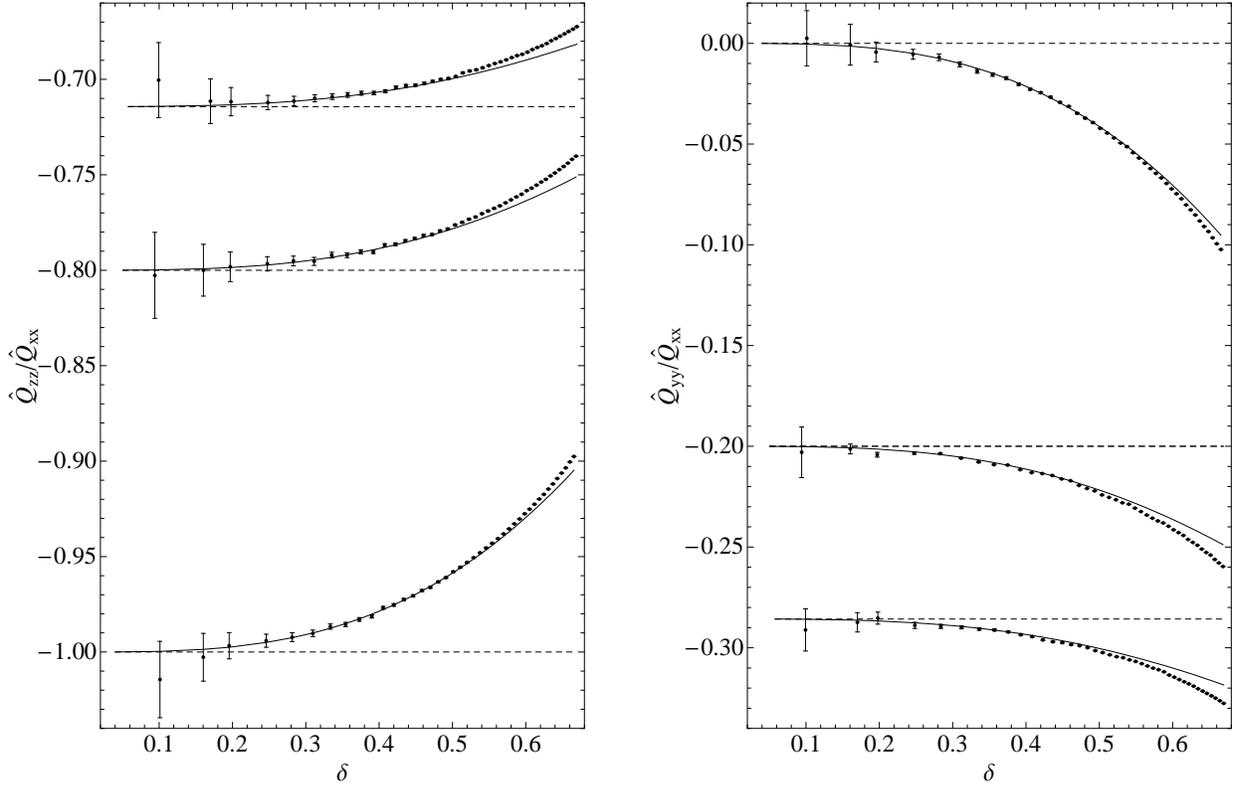} \vspace{0.3cm} 
\caption{Ratio of two
pairs of quadrupole moments for the second-order models with
$\Psi=5$, $0.1 \le \delta \le \delta_{cr}(\Psi)$, and $\nu=1,2,3$
(in the left panel, the sequence of models with different $\nu$
values runs from bottom to top; in the right panel, it runs from
top to bottom). The values obtained from numerical integration
over the entire triaxial volume (dots) are compared to the
analytical approximations (solid line) given by
Eqs.~(\ref{Qzz2/Qxx2}) and (\ref{Qyy2/Qxx2}); the analytical
estimates of the ratios for first order models are also shown
(dashed horizontal lines). The propagation of the errors of the
numerical integration leads to the plotted error bars.}\vspace{0.2cm}
\label{quadrup}
\end{figure*}

\subsection{Global quantities}
 
The previous discussion has focused on the shape of the isodensity
surfaces of the models. In particular, some interesting
conclusions have been derived based on a local analysis of the
central region and of the outer boundary of the configuration. The
maximal departures from spherical symmetry are reached at the
periphery, but these hinge on the distribution of the very small
number of stars that populate the outer region of the cluster. We
may thus wish to study some global quantities that better
characterize whether significant amounts of mass (and,
correspondingly, of light) are actually involved in the deviation
of the model from spherical symmetry. One standard such global
measure is provided by the quadrupole moment tensor
\begin{eqnarray}\label{quad}
&\displaystyle  Q_{ij}=\int_V\,(3x_ix_j-r^2\delta_{ij})\rho({\bf r}) d^3r=\nonumber\\
& \displaystyle \hat{A}r_0^5\int_V\,(3\hat{x}_i\hat{x}_j-\hat{r}^2\delta_{ij})
\hat{\rho}({\bf \hat{r}}) d^3\hat{r}=\hat{A}r_0^5\hat{Q}_{ij}~,&
\end{eqnarray}
with the integration to be performed in the volume $V$ of the
entire configuration. Here the notation for the function
$\hat{\rho}$ and for the constant $\hat{A}$ is the same as in
Eq.~(11) of Paper I. In contrast with the frequently used inertia
tensor $I_{ij} = \int_V \rho x_i x_j d^3r$ \citep[e.g., see][chap.
2]{Cha69}, the quadrupole moment is defined in such a way that in
the spherical limit it vanishes identically. In our coordinate
system it is diagonal. Note that the tidal distortions require
that the non-vanishing terms of the inertia tensor follow the
ordering $I_{xx} \geq I_{yy} \geq I_{zz}$; to visualize the
geometry of the system, we may thus also refer to the average
polar and equatorial eccentricities $\bar{e}$ and $\bar{\eta}$
defined by the relations $I_{yy} = (1 - {\bar{\eta}}^2)I_{xx}$,
$I_{zz} = (1 - {\bar{e}}^2)I_{xx}$. In general, we have
$Q_{yy}/Q_{xx}=
({\bar{e}}^2-2{\bar{\eta}}^2)/({\bar{e}}^2+{\bar{\eta}}^2)$ and
$Q_{zz}/Q_{xx}=
({\bar{\eta}}^2-2{\bar{e}}^2)/({\bar{e}}^2+{\bar{\eta}}^2)$, with
the prolate configuration identified by ${\bar{e}} =
{\bar{\eta}}$, i.e. $Q_{yy}/Q_{xx} = Q_{zz}/Q_{xx} = - 1/2$.
 
Since most of the mass is contained in the inner regions, global
quantities can be evaluated approximately by neglecting the
contribution from the region corresponding to the boundary layer.
We can thus use the second-order solutions for $\rho$ obtained in
Paper I by the method of matched asymptotic expansions and
conveniently reduce the calculation of global quantities to an
easier integration in spherical coordinates inside the sphere of radius
$\hat{r}_{tr}$. Therefore, for the quadrupole moment tensor we
find
\begin{equation}\label{Qij2}
\hat{Q}_{ij}^{(2)}=\hat{Q}_{ij,1}\epsilon+\hat{Q}_{ij,2}\frac{\epsilon^2}{2}~.
\end{equation}
We emphasize that this estimate is expected to be a good
approximation only for those models for which the contribution  of
the boundary layer is negligible with respect to the one of the
internal sphere of radius $\hat{r}_{tr}$.
 
The relevant components on the diagonal can be expressed in terms
of the matching constants of the external solution (for the
relevant definitions, see Eqs.~(63) and (69) in Paper I)
\begin{eqnarray}\label{Qxx2}
&\displaystyle \hat{Q}_{xx}^{(2)}=\frac{2}{9}\sqrt{5\pi}
\hat{\rho}(\Psi)\left[\left(a_{20}\epsilon+
b_{20}\frac{\epsilon^2}{2}\right)\right.& \nonumber \\ 
&\displaystyle\left. -\sqrt{3}\left(a_{22}\epsilon+b_{22}\frac{\epsilon^2}{2}\right)\right]~,&
\end{eqnarray}
\begin{eqnarray}\label{Qyy2}
&\displaystyle \hat{Q}_{yy}^{(2)}=\frac{2}{9}\sqrt{5\pi}
\hat{\rho}(\Psi)\left[\left(a_{20}\epsilon+b_{20}\frac{\epsilon^2}{2}\right)\right. \nonumber \\
&\displaystyle \left.  +\sqrt{3}\left(a_{22}\epsilon+b_{22}\frac{\epsilon^2}{2}\right)\right]~,&
\end{eqnarray}
\begin{equation}\label{Qzz2}
\hat{Q}_{zz}^{(2)}=-\frac{4}{9}\sqrt{5\pi}
\hat{\rho}(\Psi)\left(a_{20}\epsilon+b_{20}\frac{\epsilon^2}{2}\right)~.
\end{equation}

We recall that the constants $a_{20}$ and $b_{20}$ are positive,
while $a_{22}$ and $b_{22}$ are negative (and larger in
magnitude). Therefore, $\hat{Q}_{xx}^{(2)}$ is positive and
$\hat{Q}_{yy}^{(2)}$ and $\hat{Q}_{zz}^{(2)}$ are negative,
consistent with the detailed elongation and compressions observed
in the density profile. A summary of the derivation of these
formulae is provided in Appendix B.
 
As a measure of the degree of triaxiality of a given
configuration, we have calculated the following ratios
\begin{equation}\label{Qyy2/Qxx2}
\displaystyle  \frac{\hat{Q}_{yy}^{(2)}}{\hat{Q}_{xx}^{(2)}}=\frac{\left(a_{20}
+b_{20}\epsilon/2 \right) +\sqrt{3}\left(a_{22}+b_{22}\epsilon/2 \right)}
{\left(a_{20}+b_{20}\epsilon/2 \right)-\sqrt{3}\left(a_{22}+b_{22}\epsilon/2 \right)}~,
\end{equation}
\begin{equation}\label{Qzz2/Qxx2}
\frac{\hat{Q}_{zz}^{(2)}}{\hat{Q}_{xx}^{(2)}}=\frac{-2\left(a_{20}+b_{20}\epsilon/2 \right) }{\left(a_{20}+b_{20}
\epsilon/2 \right) -\sqrt{3}\left(a_{22}+b_{22}\epsilon/2 \right)}~,
\end{equation}
which depend explicitly on the tidal strength parameter and
implicitly on the concentration parameter. In the limit of
vanishing tidal strength, we find
\begin{equation}\label{Qyy1/Qxx1}
\frac{\hat{Q}_{yy}}{\hat{Q}_{xx}} \sim
\frac{\hat{Q}_{yy}^{(1)}}{\hat{Q}_{xx}^{(1)}}=\frac{T_{20}(\hat{r}_{tr})
+\sqrt{3}T_{22}(\hat{r}_{tr})}{T_{20}(\hat{r}_{tr})-\sqrt{3}T_{22}(\hat{r}_{tr})}
=-\frac{\nu-1}{2\nu+1}~,
\end{equation}
\begin{equation}\label{Qzz1/Qxx1}
\frac{\hat{Q}_{zz}}{\hat{Q}_{xx}} \sim
\frac{\hat{Q}_{zz}^{(1)}}{\hat{Q}_{xx}^{(1)}}=\frac{-2T_{20}(\hat{r}_{tr})}
{T_{20}(\hat{r}_{tr})-\sqrt{3}T_{22}(\hat{r}_{tr})}=-\frac{2+\nu}{2\nu+1}~,
\end{equation}
where $T_{2m}(\hat{r})$ are the quadrupole coefficients of the tidal potential
(see Eqs.~(38) and (39) in Paper I).
This result is nontrivial, because, in this limit, numerator and
denominator are both expected to vanish. Note that only for
$\nu=1$ the ratio ${\hat{Q}_{yy}}/{\hat{Q}_{xx}} =
\mathcal{O}(\epsilon)$.
 
Earlier in this paper we mentioned that two physical scales, such
as total mass and central velocity dispersion, correspond to the
two dimensional constants $A$ and $a$ that appear in the
distribution function $f_K(H)$. In fact, the total mass of the
system is given by
\begin{equation}\label{mass}
M=\int_V\,\rho({  r}) d^3r=\hat{A}r_0^3\int_V\, \hat{\rho}({ 
\hat{r}}) d^3\hat{r}=\hat{A}r_0^3\hat{M}~.
\end{equation}
If we insert the second-order solution for $\rho$ obtained in
Paper I, we find
\begin{eqnarray}\label{M2}
& \displaystyle \hat{M}^{(2)}=\int_0^{2\pi} d\phi\int_0^{\pi}d\theta \sin \theta
\int_0^{\hat{r}_{tr}}d\hat{r}\,\hat{r}^2\,
\hat{\rho}^{(2)}(\hat{r},\theta,\phi)\nonumber \\
& \displaystyle =\hat{M}_0+\epsilon\hat{M}_1+\frac{\epsilon^2}{2}\hat{M}_2~,&
\end{eqnarray}
with
\begin{equation}\label{Mi}
\hat{M}_i= \hat{M}_i (\Psi)= -\frac{4\pi
\hat{\rho}(\Psi)}{9}\lambda_i~.
\end{equation}
Here each term of the expansion is related to the corresponding
constant with $l=0$ (i.e., the monopole term) of the expression of
the external solution of the Poisson-Laplace equation calculated
by means of the method of matched asymptotic expansion (for the
relevant definitions, see Eqs.~(59),(61),(67) in Paper I).

The quality of the analytical estimates for the total mass and the
quadrupole moment tensor has been checked by comparing the values
obtained from asymptotic analysis with those resulting from direct
numerical integration of Eq.~(\ref{mass}) and Eq.~(\ref{quad})
respectively, in which the density profile
$\hat{\rho}=\hat{\rho}[\psi({  \hat{r}})]$ is used without any
additional expansion. The integration {  of the distribution function 
over the entire space, required by those global quantities,} has been 
performed by means of an Adaptive Monte Carlo method \citep[the algorithm VEGAS,
see][\S~7.8]{NRec}, well suited for our geometry. For the quadrupole, 
the results are illustrated in Fig.~\ref{quadrup}. For the mass, the 
dimensionless function $\hat{M}^{(2)}(\Psi)$ is basically unchanged 
(within 0.5 percent) with respect to the function characterizing 
the spherical King models; the Monte Carlo integration is very accurate, 
with relative errors around $10^{-5}$, and the analytical approximation 
given by Eq.~(\ref{M2}) shows an excellent agreement for every value of 
$\Psi$ in the whole range of the extension parameter $[0,\delta_{cr}(\Psi)]$.
  
\begin{figure}[t!]
\epsscale{1.15}\plotone{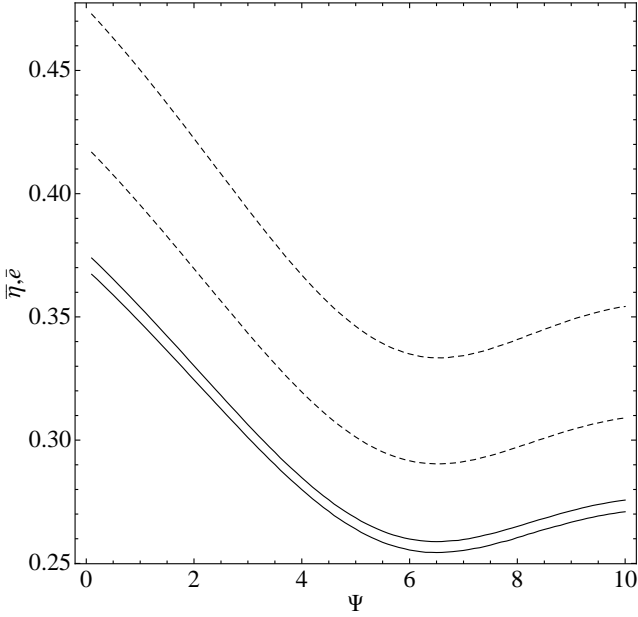} 
\caption{Average
polar ($\bar{e}$; dashed) and equatorial ($\bar{\eta}$; solid) eccentricities
for critical second-order models as a function of concentration,
with $0.1 \le \Psi \le 10$, for two different host potentials.
Models with $\nu=3$ correspond to the inner pair of curves, while
models with $\nu=1$ to the outer pair. The eccentricities have
been determined numerically from their definitions (see main text)
in terms of $\hat{\rho}=\hat{\rho}[\psi({\bf \hat{r}})]$, without
any additional expansion; the relative errors are around $10^{-5}$.}
\label{ecceglob}
\end{figure}
  
The average eccentricities for critical second-order models as a
function of concentration, with $0.1 \le \Psi \le 10$, for two
different choices of the host galaxy potential ($\nu=3,1$), are
shown in Fig.~\ref{ecceglob}. The values are calculated directly
from the definitions given earlier in this subsection in terms of
$\hat{\rho}=\hat{\rho}[\psi({  \hat{r}})]$ with no additional
expansions. A non-monotonic dependence on concentration is
revealed, with generally higher average eccentricities attained by
low-concentration models. The trends of the polar and equatorial
eccentricities are basically the same, as shown by the fact that
the related curves in the plot can be matched approximately by a
rigid translation. As expected (see \S~3.1), models with $\nu=1$
show a larger separation between polar and equatorial
eccentricities than models with $\nu=3$. The presence of a minimum
for the curves at $\Psi\approx 6.5$ occurs approximately at the
location where the function $Log[\epsilon_{cr}(\Psi)]$ shows an inflection 
point (regardless of the value of $\nu$; see Fig.~\ref{parsp}). 
 
\begin{figure*}[t!]
\epsscale{1.15}\plotone{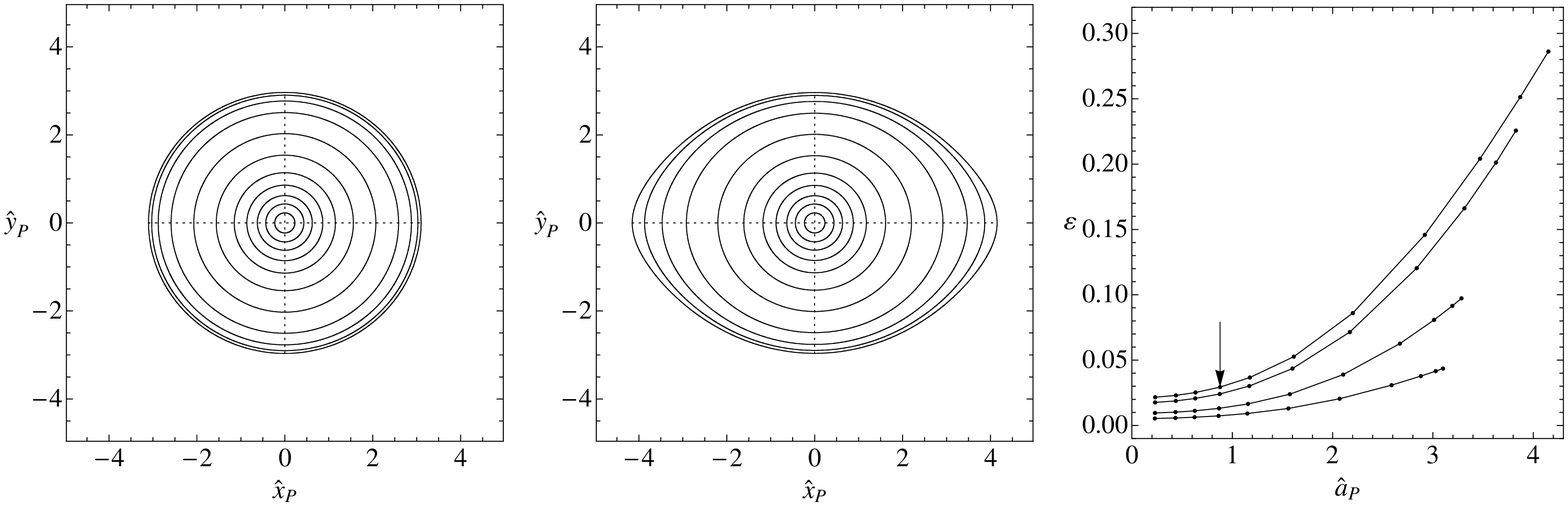} \caption{Projections of
a second-order critical model ($\Psi=2$ and $\nu=3$) along the
$\hat{x}$ axis (left panel [a]) and along the $\hat{y}$ axis
(central panel [b]). The ellipticity profiles (right panel [c],
from bottom to top) refer to the projections along directions
identified by $(\theta=\pi/2,\phi=i\, \pi/6)$ with $i=0,...,3$;
dots represent the locations of the isophotes drawn in panels
[a],[b], which correspond to selected values of $\Sigma/\Sigma_0$
in the range $[0.9,10^{-6}]$. The arrow indicates the position of
the half-light isophote (practically the same for every projection
considered in the figure).}\vspace{0.2cm} \label{ecceproj}
\end{figure*}

\subsection{Projected density profile}
 
Under the assumption of a constant mass-to-light ratio, projected
models can be compared with the observations.  We have then
computed surface (projected) density profiles and projected
isophotes. The projection has been performed along selected
directions, identified by the viewing angle $(\theta,\phi)$,
corresponding to the $\hat{z}_P$ axis of a new coordinate system
related to the intrinsic system by the transformation ${ 
\hat{x}}_P=R{  \hat{x}}$; the rotation matrix
$R=R_1(\theta)R_3(\phi)$ is expressed in terms of the viewing
angles, by taking the $\hat{x}_P$ axis as the line of nodes
\citep[see][for an equivalent projection rule]{Ryd91}. Given the
symmetry of our models (see \S~3.1), viewing angles can be chosen
from the first octant only. In particular, we used a $4 \times 4$
polar grid defined by $\theta_i=i\,\pi/6$ and $\phi_j=j\,\pi/6$
with $i,j=0,..,3$, and calculated (by numerical integration, using
the Simpson rule) the dimensionless projected density
\begin{equation}
\hat{\Sigma}(\hat{x}_P,\hat{y}_P)=\int_{-\hat{z}_{sp}}^{\hat{z}_{sp}}
\hat{\rho}({  \hat{r}}_P)d\hat{z}_P~,
\end{equation}
where $\hat{z}_{sp}=(\hat{x}_e^2-\hat{x}_P^2-\hat{y}_P^2)^{1/2}$
with $\hat{x}_e$ the edge of the cluster along the $\hat{x}$ axis
of the intrinsic coordinate system (i.e., we ``embedded" the
triaxial configuration in a sphere of radius given by its maximal
extension). The projection plane $(\hat{x}_P,\hat{y}_P)$ has been
sampled on an equally-spaced $108 \times 108$ square grid centered
at the origin (note that for
$\hat{x}_P^2+\hat{y}_P^2\ge\hat{x}_e^2$ the projected density is
correctly set to zero).
 
The morphology of the isophotes of a given projected image can be
described in terms of the {\it ellipticity} profile, defined as
$\varepsilon=1-\hat{b}_P/\hat{a}_P$ where $\hat{a}_P$ and
$\hat{b}_P$ are the semi-axes, as a function of the major axis
$\hat{a}_P$. As already noted for the (intrinsic) eccentricity
profiles, the deviation from circularity increases with the
distance from the origin. In the inner region, the ellipticity is
consistent with the central eccentricities $e_0$ and $\eta_0$
calculated in \S~3.2.
 
By taking lines of sight different from the axes of the symmetry planes, 
we have also checked whether the projected models would exhibit isophotal
twisting. For all the cases considered, the position angle of the
major axis remains unchanged over the entire projected image.
Tests made by changing the resolution of the grid confirm that, even in 
the most triaxial case ($\nu=1$), no significant twisting is present.
 
The first two panels of Fig. \ref{ecceproj} show the projected
images of a critical second-order model with $\Psi=2$ and $\nu=3$
along the $(\pi/2,0)$ and $(\pi/2,\pi/2)$ directions, (i.e., the
$\hat{x}$ and $\hat{y}$ axis of the intrinsic system),
corresponding, respectively, to the least and to the most
favorable line of sight for the detection of the intrinsic
flattening of the model. For the same model, the third panel
illustrates the ellipticity profiles for various lines of sight.

Figure \ref{densproj} shows the surface density profiles along the
$\hat{x}_P$ and $\hat{y}_P$ axes of the projection plane for ten
critical second-order models with $\nu=3$, viewed along the
$(\pi/2,\pi/2)$ direction. As a further characterization, for the
same models in the lower panel we also present the surface density
profiles obtained by averaging the projected density distribution
on circular annuli; this conforms to the procedure often adopted
by observers in dealing with density distributions with very small
departures from circular symmetry \citep[e.g., see][]{Lan07}. As expected, 
circular-averaged profiles lie between the corresponding regular profiles 
taken along the principal axes of the projected image.

\section{Intrinsic and projected kinematics}
 
By construction, the models are characterized by isotropic
velocity dispersion. The intrinsic velocity dispersion can be
determined directly as the second moment in velocity space
(normalized to the intrinsic density) of the distribution function
\begin{equation}
\sigma^2(\psi)=\frac{2}{5a}\frac{\gamma\left(7/2,\psi \right)}
{\gamma\left(5/2,\psi\right)}=\frac{1}{a}\hat{\sigma}^2(\psi)~,
\end{equation}
where $\gamma$ represents the incomplete gamma function (near the
boundary of the configuration, the velocity dispersion profile
scales as $\hat{\sigma}^2(\psi)\sim (2/7) \psi$). This shows that
the isodensity surfaces of the models are in a one-to-one
correspondence with the isovelocity and isobaric surfaces (defined
by $\sigma^2[\psi({  \hat{r}})]=$ const). As noted for the
intrinsic density profiles in \S~3.1, a compression along the
$\hat{z}$ axis and an elongation along $\hat{x}$ axis occur also
for the intrinsic velocity dispersion profiles.
In Fig.~\ref{velproj}.a we present the intrinsic velocity dispersion 
profiles along the $\hat{x}$ axis for the same critical models illustrated 
in Fig.~\ref{dens} compared to the profiles of the corresponding spherical King
models. The behavior of the projected velocity dispersion profiles
near the boundary is significantly different from that of the spherical 
models. 

\begin{figure}[t!]
\epsscale{1.22} \plotone{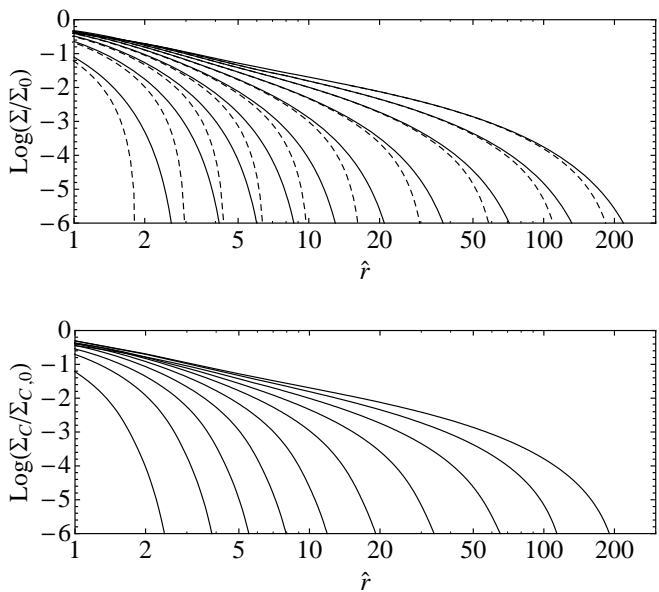} \caption{Projected
density profiles (normalized to the central value) for the same
ten second-order critical models displayed in Fig.~\ref{dens}. Top
panel (a): the models are viewed from the $\hat{y}$ axis and the
profiles taken along the two principal axes in the projection plane
(along $\hat{x}_P$ (solid) and $\hat{y}_P$ (dashed), i.e. along 
the $\hat{x}$ and $\hat{z}$ axes of the intrinsic frame of
reference).
Bottom panel (b): the projection is performed on the same line of
sight of the previous panel, but the profiles are taken by
averaging the projected surface density on circular annuli, as if
the image were intrinsically circular.} \label{densproj}
\end{figure}

\begin{figure}[t!]
\epsscale{1.22} \plotone{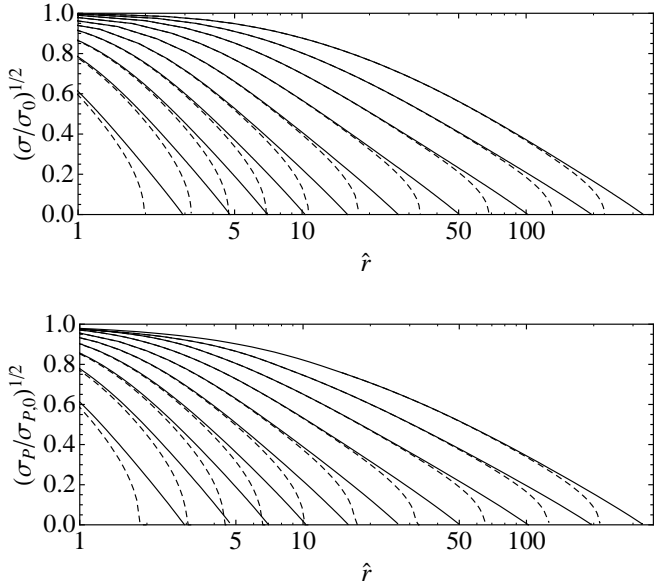} \caption{Top panel (a): 
intrinsic velocity dispersion profiles (solid, normalized to the central value)
along the $\hat{x}$ axis for the same ten second-order critical models 
illustrated in Fig.~\ref{dens}.a compared to the corresponding spherical King models (dashed).
Bottom panel (b): projected
velocity dispersion profiles (normalized to the central value) for
the same ten second-order critical models illustrated in
Fig.~\ref{densproj}.a, viewed along the same direction. Solid and dashed
lines show the profiles along the $\hat{x}_P$ axis and $\hat{y}_P$ of the
projection plane, respectively.} \label{velproj}
\end{figure}

The projected velocity moments can be calculated by integrating
along the line of sight (weighted by the intrinsic density) the
corresponding intrinsic quantities. Therefore, the projected
velocity dispersion is given by
\begin{eqnarray}
& \displaystyle \sigma^2_{P}(\hat{x}_P,\hat{y}_P)=\frac{\int_{-\hat{z}_{sp}}^{\hat{z}_{sp}}
\sigma^2({\bf \hat{r}}_P) \rho({\bf \hat{r}}_P)d \hat{z}_P}
{\Sigma(\hat{x}_P,\hat{y}_P)}=&\nonumber\\
& \displaystyle \frac{2}{5a}\frac{\int_{-\hat{z}_{sp}}
^{\hat{z}_{sp}}\gamma[7/2,\psi({\bf \hat{r}}_P)]\exp[\psi({\bf
\hat{r}}_P)]
d\hat{z}_P}{\hat{\Sigma}(\hat{x}_P,\hat{y}_P)}=\frac{1}{a}
\hat{\sigma}^2_{P}(\hat{x}_P,\hat{y}_P)~.&
\end{eqnarray}
Figure ~\ref{velproj}.b shows the projected velocity dispersion profiles 
along the $\hat{x}_P$ and $\hat{y}_P$ axis of the projection plane for 
the same models displayed in Fig.~\ref{densproj} (the line of sight is 
defined by $(\pi/2,\pi/2)$).

\section{Discussions and conclusions}

\subsection{A comment on the complex structure of the outer regions}

In Paper I we have noted the singular character of the mathematical 
problem associated with the free boundary set by the three-dimensional 
surface of these tidally truncated models. In \S~2 of the present Paper, 
we have further emphasized the additional singularity that characterizes 
close-to-critical models (see comment after Eq.~(\ref{eps0}), which prompted the 
analysis described in the Appendix A). As is true in general in the study 
of boundary layers and similar problems, it is no wonder that in the 
vicinity of such critical boundaries, a number of complex physical effects 
may take place and play an important role in determining the detailed 
structure of the solution. On the other hand, the properties of the derived 
solution away from the boundary are quite robust (see \S~3.3 and the 
Appendix). As to some of the subtle properties of the expected distribution 
function close to the edge of a cluster, it is interesting to summarize 
here below the main results that emerge from a vast body of evolutionary 
models (N-body, Fokker-Planck, Monte Carlo, gas), on the issue of the 
interplay between pressure anisotropy and tidal effects.         

Since the first solutions of the Fokker-Planck equation by means of a 
Monte Carlo approach, as described by \citet{Hen71} or \citet{SpiHar71}, 
it has been shown that one-component {\em isolated spherical} clusters, 
starting from a variety of initial conditions \citep[see][chap.4 and 
references therein]{Spi87}, develop a core-halo structure in which the 
central parts are almost isotropic while the outer regions are 
characterized by {\em radial} anisotropy. A commonly reported interpretation 
is that the halo is mainly populated by stars scattered out from the core 
on radial orbits. 
If the evolution of a cluster is influenced not only by internal processes
but also by the presence of an external tidal field, the growth of pressure 
anisotropy is significantly damped. Direct N-body simulations \citep[e.g., 
see][]{GieHeg97,AarHeg98,BauMak03,LeeLeeSun06}, anisotropic Fokker-Planck 
\citep[e.g., see][]{TakLeeIna97}, and Monte Carlo 
models \citep[e.g., see][]{Gie01} of both single and multi-mass systems 
suggest that clusters in circular orbits \citep[and even in eccentric 
orbits, see][]{BauMak03}, starting from isotropic initial conditions, 
tend to preserve pressure isotropy, except for the outermost parts which 
become {\em tangentially} anisotropic due to the preferential 
loss of stars on radial orbits, induced by the tidal field.  The overall 
agreement on this result is nontrivial, because of the aforementioned 
differences in the treatment of the external tidal field. 
Even extreme cases of time-dependent tides, such as disk shocking, influence
the degree of pressure anisotropy since it has been shown that they may 
represent a dominant mechanism (``shock relaxation") of the energy 
redistribution, leading to a substantial isotropy, of the weakly bound 
stars \citep[see][both papers are based on a Fokker-Planck approach]{OhLin92,Kun95}.    

These results confirm that, of course, the properties of the models 
constructed in Paper I and described in this paper should be taken only 
as a zeroth-order reference frame, to single out the precise role of 
external tides, and should not be taken literally as a realistic 
representation of real objects since a number of simplifying assumptions 
are made. On the other hand, by comparing data with 
such an idealized reference model it will be possible to better assess 
the role of tides with respect to other physical ingredients studied 
separately. 

\subsection{Summary and concluding remarks}

The main results that we have obtained from the detailed analysis of 
the family of tidal triaxial models can be summarized as follows:
\begin{enumerate}
\item Two tidal regimes exist, namely of low and high-deformation, 
which are determined by the combined effect of the tidal strength of 
the field and of the concentration of the cluster. The degree 
of deformation increases with the degree of filling of the relevant 
Roche volume. Far from the condition of Roche volume filling, the models 
are almost indistinguishable from the corresponding spherical King 
models. A number of studies have investigated the evolution 
of tidally perturbed stellar systems initially underfilling their Roche 
lobe \citep[e.g., see][]{GieBau08,Ves09}, concluding that 
some of the relevant dynamical processes, in particular evaporation, depend 
on the degree of filling of the Roche volume.   
The intrinsic properties of the models discussed in this paper can be 
useful for setting self-consistent nonspherical initial conditions of  
numerical simulations aiming at studying in further details the evolution of 
configurations starting from sub-critical tidal equilibria.
\item For a given tidal strength, there exists a maximum value of 
concentration for which a closed configuration is allowed (see 
Fig.~\ref{parsp}). In such ``critical" case, the truncation radius of 
the corresponding spherical King models is of the same order of the 
tidal radius of the triaxial model. The shape of the boundary of the 
maximally deformed models is given by the ``critical" zero-velocity 
surface of the relevant Jacobi integral and is basically independent of 
concentration, while the deformation of the internal region strongly 
depends on the value of the critical tidal strength and is more 
significant for low-concentration models. This statement agrees with a
general trend noted by \citet{WhiSha87} for the globular cluster system 
of our Galaxy.  
\item The structure of the models can be described in terms of the polar 
and equatorial eccentricity profiles of the intrinsic isodensity surfaces. 
The maximal departures from spherical symmetry are reached in the outskirts. 
\item For finite tidal strength, the central values of the polar and 
equatorial eccentricities are finite, $\mathcal{O}(\epsilon^{1/2})$; 
this result is nontrivial since the tidal potential which induces 
the perturbation is a quadratic homogeneous function of the spatial 
coordinates.  
\item Global measures of the degree of triaxiality in terms of the 
quadrupole moment tensor have been introduced and calculated for different 
values of the tidal strength and different potentials of the host galaxy.
The potential of the host galaxy sets the geometry of 
the tidal perturbation, as nicely shown by the analytic expressions 
for the ratios of the components of the quadrupole moment tensor, which, 
in the case of first-order models, reduce to simple functions of the 
coefficient $\nu$.    
Average eccentricities have been calculated from the inertia tensor 
components, evaluated numerically.   
\item As a key step in the direction of a comparison with observations, 
projected density profiles and ellipticity profiles have 
been calculated for a number of models for several lines of sight. 
\item The study of the relevant (projected) isophotes indicates that 
no isophotal twisting occurs. This result is nontrivial since the 
models are nonstratified and centrally-concentrated, conditions 
under which, in principle, isophotal twist may occur (see \citet{Sta77}; 
models based on St\"{a}ckel potentials are also known to be twist-free, 
as shown by \citet{Fra88}). 
\item Finally, close to the boundary, the intrinsic and projected 
kinematics shows significant differences with respect to that of 
spherical models. 
\end{enumerate}
 
Since our models are all characterized by monotonically increasing 
ellipticity profiles, they cannot explain the variety of behavior 
of observed ellipticity profiles \citep[see][]{Gey83}, but the range 
of the predicted flattening ($\varepsilon < 0.3$) is consistent with that 
observed in most globular clusters (see \citet{WhiSha87} for the clusters in 
the Milky Way and \citet{FreFal82} for those in the LMC). 
Therefore, with the exception of special clusters such as Omega Centauri,
we think that the modest but frequently observed deviations from 
spherical symmetry might have their origin traced to tides. 

Finally, since our models are intrinsically more elongated (in the direction of 
the center of the host galaxy) than spherical King models, they might be useful 
for interpreting clusters with a surface brightness profile extending 
beyond that predicted by the spherical King models. Recent investigations 
\citep[see][]{McLvdM05} suggest that such ``extra-tidal" structures are a 
fairly generic feature, especially for extragalactic clusters, and not 
just a transient property, present only at young ages. 
   
On the basis of the work presented here, in a following paper we plan to address in 
detail the issues raised by observations.

\acknowledgments

We wish to thank D.~C. Heggie for several discussions and for his code, 
available within the software enviroment STARLAB by P. Hut, 
S. McMillan, J. Makino, and S. Portegies Zwart. We also thank D. Chernoff,
L. Ciotti, and E. Vesperini for a number of useful comments and suggestions. 
Our code implementing the iteration method, described in Appendix A,
makes use of the package S2kit 1.0 by P.J. Kostelec and D.N. 
Rockmore, and of FFTW 3.2.1 by M. Frigo and S.~G. Johnson. 
The final part of this work was carried out at The Kavli Institute for 
Theoretical Physics, Santa Barbara, while we participated in the Program 
``Formation and Evolution of Globular Clusters".

\appendix

\section{Perturbation vs. iteration}
 
For completeness, we calculated selected models also by means of the 
iteration method described in \S~5.2 of Paper I, 
in order to verify the quality of the solution obtained with the method of 
matched asymptotic expansions, in particular in the critical regime.
This technique follows the approach proposed by \citet{PreTom70}. The 
basic idea is to get an improved solution of the Poisson equation (see 
Eq.~(77) in Paper I) by evaluating the right-hand side with the solution 
obtained from the immediately previous step, starting from the spherical 
King models taken as initial ``seed solutions".
In our code, the required spherical harmonic analysis and synthesis of density 
and potential have been performed by means of the S2kit 1.0 
package \citep{KosRoc04}, which makes use of FFTW 3.2.1 \citep{FriJoh05}. 
We decided to truncate the harmonic series at 
$l=4$ in order to be consistent with the maximum harmonic index admitted 
by the second-order asymptotic solution. The iteration stops when 
convergence to four significant digits in the whole domain of the solution 
is reached.  

For selected values of the concentration parameter in the range $0.1\le \Psi \le 10$ 
(for simplicity, we considered only the case of an external potential with 
$\nu=3$), the corresponding critical value of the tidal strength parameter 
obtained with the iteration method is consistent to $10^{-3}$ with the value determined 
by the numerical solution of Eq.~(\ref{sistcrit1}), in which the constants 
obtained from the asymptotic matching are used. For a critical model up 
to 20 iteration steps are required for convergence, while a subcritical 
solution typically takes in 4 to 8 steps.       
    
\section{Global quantities from the multipole expansion of the cluster
potential}
 
Based on the expansion in spherical harmonics of $1/|{ 
\hat{r}}-{  \hat{r}'}|$ given in Eq.~(3.70) of \citet{Jac99},
the external potential generated by our model can be expressed
by means of the following multipole expansion
\begin{equation}\label{Green}
a\Phi_C^{\mbox{\tiny (ext)}}({  \hat{r}})=-\frac{9}{4\pi\hat{\rho}(\Psi)}\int_Vd^3\hat{r}'
\frac{\hat{\rho}({  \hat{r}'})}{|{  \hat{r}}-{  \hat{r}'}|}=
-\frac{9}{\hat{\rho}(\Psi)}\sum_{l=0}^{+\infty}\frac{1}{2l+1}
\sum_{m=-l}^{+l}\hat{q}_{lm}\frac{Y_{lm}(\theta,\phi)}{\hat{r}^{l+1}}~,
\end{equation}
where the multipole coefficients are defined as
\begin{equation}\label{qlm}
\hat{q}_{lm}=\int_V d^3\hat{r}'Y_{lm}(\theta',\phi')\hat{r}'^{\,l}
\hat{\rho}({  \hat{r}'})~,
\end{equation}
with the integration to be performed in the volume $V$ with
boundary surface defined by $\psi = 0$. This general expression
can be compared with the second-order solution of the Laplace
equation obtained in Paper I
\begin{eqnarray}\label{multip}
\left(a\Phi_C^{\mbox{\tiny (ext)}}\right)^{(2)}({  \hat{r}})&=&\left[ \lambda_0+
\lambda_1\epsilon+\lambda_2\frac{\epsilon^2}{2}\right]\frac{1}{\hat{r}}
+ \left[a_{20}\epsilon + b_{20}\frac{\epsilon^2}{2}\right]
\frac{Y_{20}(\theta,\phi)}{\hat{r}^3}+\left[a_{22}\epsilon+b_{22}
\frac{\epsilon^2}{2}\right]\frac{Y_{22}(\theta,\phi)}{\hat{r}^3}
\nonumber\\
&&+b_{40}\frac{\epsilon^2}{2}\frac{Y_{40}(\theta,\phi)}{\hat{r}^5}+
b_{42}\frac{\epsilon^2}{2}\frac{Y_{42}(\theta,\phi)}{\hat{r}^5}+
b_{44}\frac{\epsilon^2}{2}\frac{Y_{44}(\theta,\phi)}{\hat{r}^5}~,
\end{eqnarray}
in order to determine the relation between the second-order multipole 
coefficients $\hat{q}_{lm}^{(2)}$ (i.e. calculated by means of the 
second-order expression for the density) and the matching constants 
that appear in Eq.~(\ref{multip}). 
Therefore, we find that
\begin{equation}\label{zero}
\lambda_0+\lambda_1\epsilon+\lambda_2\frac{\epsilon^2}{2}=
-\frac{9}{\sqrt{4\pi}\hat{\rho}(\Psi)}\,\hat{q}_{00}^{(2)}~,
\end{equation}
so that Eq.~(\ref{Mi}) follows. For the higher-order harmonics we
find the following relations for the non-vanishing coefficients
\begin{equation}\label{q2m}
a_{2m}\epsilon+b_{2m}\frac{\epsilon^2}{2}=-\frac{9}{5\hat{\rho}(\Psi)}
\,\hat{q}_{2m}^{(2)}~,
\end{equation}
for $m=0,2$ and
\begin{equation}\label{q4m}
b_{4m}\frac{\epsilon^2}{2}=-\frac{1}{\hat{\rho}(\Psi)}\,\hat{q}_{4m}^{(2)}~,
\end{equation}
for $m=0,2,4$.
 
Recalling that we are using {\em real} spherical harmonics with
the Condon-Shortley phase, we get
\begin{equation}\label{q20}
\hat{q}_{20}=\frac{1}{4}\sqrt{\frac{5}{\pi}}\hat{Q}_{zz}~,
\end{equation}
\begin{equation}\label{q22}
\hat{q}_{22}=\frac{1}{12}\sqrt{\frac{15}{\pi}}(\hat{Q}_{xx}-\hat{Q}_{yy})~.
\end{equation}
To determine all the nontrivial components of the quadrupole
moment tensor, we use  the condition
$\mbox{Tr}(\hat{Q}_{ij})=0$. Therefore, for the second-order
solution of Paper I, the system
\begin{equation}\label{sistquad}
\left\{
\begin{array}{lll}
\displaystyle
\hat{Q}_{xx}^{(2)}-\hat{Q}_{yy}^{(2)}=-\frac{20}{3}\sqrt{\frac{\pi}{15}}
\hat{\rho}(\Psi)\left( a_{22}\epsilon+b_{22}
\frac{\epsilon^2}{2}\right) \vspace{.1cm}\\
\displaystyle
\hat{Q}_{zz}^{(2)}=-\frac{20}{9}\sqrt{\frac{\pi}{5}}\hat{\rho}
(\Psi)\left( a_{20}\epsilon+b_{20}\frac{\epsilon^2}{2}\right) \vspace{.1cm}\\
\displaystyle
\hat{Q}_{xx}^{(2)}+\hat{Q}_{yy}^{(2)}+\hat{Q}_{zz}^{(2)}=0~,
\end{array}
\right.
\end{equation}
leads to the expressions recorded in Eqs.~(\ref{Qxx2})-(\ref{Qzz2}).

\end{document}